\documentclass[pdflatex,sn-mathphys-num]{sn-jnl}

\usepackage{graphicx}%
\usepackage{multirow}%
\usepackage{amsmath,amssymb,amsfonts}%
\usepackage{amsthm}%
\usepackage{mathrsfs}%
\usepackage[title]{appendix}%
\usepackage{xcolor}%
\usepackage{textcomp}%
\usepackage{manyfoot}%
\usepackage{booktabs}%
\usepackage{algorithm}%
\usepackage{algorithmicx}%
\usepackage{algpseudocode}%
\usepackage{listings}%
\usepackage{hyperref}%

\theoremstyle{thmstyleone}%
\theoremstyle{thmstyletwo}%

\theoremstyle{thmstylethree}%
\newtheorem{definition}{Definition}%

\begin{document}

\title[e-values for multiverse analysis]{ The role of e-values in multiverse analysis }


\author[1]{\fnm{Paul} \sur{Rognon-Vael}}\email{paul.rognon@gmail.com}

\author*[2,3]{\fnm{David} \sur{Rossell}}\email{david.rossell@upf.edu}

\affil[1]{\orgdiv{BIDSA}, \orgname{Bocconi University}, \orgaddress{\street{Via Roberto Sarfatti, 25}, \city{Milano}, \postcode{20136}, \country{Italy}}}

\affil*[2]{\orgdiv{Department of Economics and Business}, \orgname{Universitat Pompeu Fabra}, \orgaddress{\street{Ram\'on Trias Fargas 25}, \city{Barcelona}, \postcode{08005}, \country{Spain}}}

\affil[3]{\orgdiv{Data Science Center}, \orgname{Barcelona School of Economics}, \orgaddress{\street{Ram\'on Trias Fargas 25}, \city{Barcelona}, \postcode{08005}, \country{Spain}}}

\abstract{Multiverse analysis refers to a situation where one wishes to assess the association between multiple possible treatment definitions and multiple possible outcome definitions, potentially within multiple sub-populations, among other possible analysis specifications. The multiverse is a useful exploratory tool to assess heterogeneity across such specifications, but it is sometimes also used to assess statistical significance. Then, it is critical to ensure the  statistical validity  of the findings. 
We study the use of e-values within generalized linear models as a  hypothesis testing tool for multiverse analysis. E-values allow one to control the false discovery rate (FDR) across multiple tests, and to flexibly combine evidence across hypotheses and across datasets. 
We compare  via simulations  the performance of several approaches: universal e-values, soft-rank e-values, and p-to-e calibration. For problem characteristics typically encountered in multiverse analyses,  a mixture-based and an inverse root based  p-to-e calibration significantly outperform the other two approaches in statistical power.
Said power is however moderate unless the sample size or effect sizes are large enough,  and slightly lower than for Benjamini-Yekutieli adjusted p-values, unless the multiverse size is large. Hence, e-values bring limited benefits purely for FDR control, their main asset being aggregating evidence over hypotheses or over datasets. We illustrate these features to  
study the association between teenager technology use and mental well-being, where we find an association between depression, low self-esteem and peer problems with internet, social media  and television use. }




\maketitle

\section{Introduction}
\label{sec:intro}

There are many applications where one wants to assess the effect of multiple treatments on multiple outcomes, potentially also separately within multiple sub-populations.
A prominent example in the Social Sciences occurs when one has multiple outcomes that measure related underlying characteristics, such as different measures of psychological well-being \citep{orben:2019}, several measures of economic activity \citep{banerjee_abhijit:2015}, or several measures of radicalization in social media usage \citep{bail:2018}.
Similarly, the multiple treatments may refer to using various types of technology, realizing various anti-poverty interventions, or exposure to various social media unradicalization campaigns.
It may also be natural to wish to assess treatment effects within sub-populations, for example defined by gender and age groups, economic regions, or social media user backgrounds.
Beyond the Social Sciences, similar situations arise in other fields. In Biomedicine one may study the effect of non-identical, but related, treatments on multiple measures of disease progression \citep{pocock:1987}, in Climate Science one may study the association between multiple human-driven processes on multiple physical parameters \citep{knutti:2010}, etc.
 Furthermore, there may also be more than one dataset that one can conceivably use. 

Multiverse analysis \citep{steegen:2016} provides a data analysis framework to consider such multiple outcome-treatment-subpopulation-dataset  combinations.
A key pitfall that it seeks to avoid is selective reporting, whereby researchers consider all 
combinations and report only those where the estimated effects are larger or more statistically significant. Critically, selective reporting occurs when a researcher fails to indicate all the combinations that were considered.
By acknowledging all such combinations, multiverse analysis serves as an excellent exploratory data analysis method to assess the heterogeneity of results across various analysis specifications.
 Further, if one observes consistent trends across comparisons (for example, similar estimated treatment effect on multiple outcomes) or across datasets, one may informally have stronger evidence that said treatment effects exist. 
In some instances, researchers may be interested in going beyond a descriptive analysis to also assess statistical significance. 
 In this paper we investigate the use of e-values as a tool to assess significance within a multiverse analysis. 
E-values are an increasingly popular hypothesis testing framework that has several features relative to p-values.
 First, e-values are well-suited to control the false discovery rate (FDR) in multiple comparisons, and in situations where one tests many hypotheses they can have higher statistical power than Benjamini-Yekutieli p-value adjustments \citep{wang_ruodu:2022,benjamini:2001}. There are also settings where the power can be lower, as we discuss here. Second, e-values enable anytime-valid inference, which allows accumulating evidence across multiple datasets and making decisions sequentially, based on the evidence accumulated at any given point in time. 
Third, e-values allow combining evidence across hypotheses, for example on the effect of a treatment across multiple outcomes, or on its effect in multiple sub-populations. In summary, e-values give the researcher a fair amount of flexibility for testing hypotheses while preserving statistical validity, and both issues are relevant for multiverse analysis. 
For seminal papers on e-values see \cite{vovk:2021,gruenwald:2024,chugg:2026}, and for an excellent monograph see \cite{ramdas:2025}.
To be clear, we do not argue that hypothesis testing should be preferred over descriptive analyses or confidence intervals, simply that if one wishes to test hypotheses then this should be done properly.

Before proceeding, we discuss a popular alternative strategy to reporting all the performed treatment-outcome-subpopulation-dataset comparisons: focus the analysis on some summary measure across all comparisons. 
This is the standard approach for meta-analyses in Biomedicine, where one combines the evidence for the effect of a medical treatment (or family of related treatments) across studies (typically, randomized clinical trials). 
Therein, it is common to account for treatment effect heterogeneity, often driven by differences in the patient population targeted by each study.
A notable example in the Social Sciences, where randomized trials are rare, is Specification Curve Analysis (SCA, \cite{simonsohn:2020}). 
 SCA is a useful descriptive tool to assess whether treatment effects are heterogeneous or vary substantially across analysis specifications, or to the contrary whether they are robust to the specification. SCA can also be used for hypothesis testing: 
by default, SCA reports the median estimated effect across treatment-outcome-subpopulations, and assesses the statistical significance of this single parameter.
An advantage of using a single summary measure over multiple comparisons is that, since no significance corrections are needed, if all treatment effects are comparable then one may have higher statistical power to detect that the summarized effect truly exists.
However, we urge strong caution in the sense that aggregate measures should only be used in appropriate circumstances.
A critical pitfall is that, when the treatments are not comparable or there is significant heterogeneity in their effects, the aggregate measure may be highly misleading.
For example, \cite{semken:2022} showed that SCA led \cite{orben:2019} to erroneously concluding that the association between technology use and teenager well-being was practically negligible. The reason was that the association was near-zero for some technologies such as television use, but large for others such as mobile phones, and hence overall mild on the average.
Another issue arises when the aggregate measure is obtained across multiple possible subsets of control covariates, an option that is explicitly considered in SCA.
This is inadequate because one combines quantities that have a different statistical meaning, such as marginal and conditional associations, and hence the summary has no natural interpretation. 
For example, assessing the conditional association between salary and gender given an individual's job position, education and other characteristics has a different meaning than assessing the marginal association between salary and gender. Namely, a marginal association between gender and salary can be partially driven by people from different genders making different career choices, whereas an association between gender and salary within people with the same job position, education and characteristics suggests a possible salary discrimination. 
See \cite{semken:2022} for an in-depth discussion of issues with SCA.

We briefly review the rich literature on FDR control for multiple testing \citep{benjamini:1995}.
The FDR is the expected proportion of false positives among the hypotheses that are claimed as statistically significant.
\cite{benjamini:1995} proposed a p-value adjustment procedure that controls the FDR under the assumption that the tests are independent,  and Benjamini-Yekutieli (BY, \cite{benjamini:2001}) a variation that allows an arbitrary dependence structure between the tests. 
A large body of subsequent work extended the FDR framework in several directions.
A prominent example are the q-values of~\cite{storey:2003}.
Analogously to p-values controlling the expected number of false discoveries, q-values control the expected positive FDR, that is the expected fraction of false discoveries, conditional on $\geq 1$ discoveries being reported.
However, q-values require estimating nuisance parameters such as the proportion of true null hypotheses, and therefore do not control the FDR exactly. Further, estimating said nuisance parameters often requires simplifying assumptions such as independence between the test statistics, which rarely hold in practice.
Another influential line of work is based on the local FDR~\citep{efron:2001}. 
The local FDR can be interpreted as the posterior probability that a given hypothesis is null, given the value of the test statistic for that hypothesis. The local FDR provides a more refined, hypothesis-specific, assessment of significance than tail-area measures such as q-values. 
Estimating the local FDR accurately requires the number of hypotheses being tested to be larger than is common in multiverse analyses (at least in the hundreds), assuming independence across tests, and making certain choices when estimating the global mixture distribution of the test statistics.
Further related work includes the optimal discovery procedure of~\cite{storey:2007}, and the Bayesian discovery procedure of~\cite{guindani:2009}. See~\cite{efron:2012} for a monograph on FDR control.

 Similar to BY-adjusted p-values,  e-values provide a finite-$n$ FDR control that does not require estimating nuisance parameters, and which holds regardless of the number of tests being performed and of their dependence structure. This is relevant in multiverse analysis, where tests cannot be assumed independent.
As mentioned, another feature of e-values is that they are composable:  they can be multiplied across independent datasets or averaged across multiple hypotheses 
to obtain valid overall measures of evidence. 
A main limitation of e-values relative to p-values, q-values and empirical Bayes is that the latter usually have a higher statistical power, provided that their assumptions hold. 
Therefore, e-values are most appropriate for datasets where one has relatively large sample sizes. Otherwise, while FDR control is ensured, one may obtain few statistically significant findings. In this paper, we consider several strategies to obtain e-values and compare their power via simulation studies that mimic common sample sizes and problem dimensions encountered in multiverse analysis within the Social Sciences.  We also compare their power to BY p-value adjustment.  Our findings suggest that simple p-to-e calibration rules, which convert a p-value into an e-value, attain higher statistical power than generic strategies such as universal e-values \citep{wasserman:2020} and soft-rank e-values \citep{koning:2023}.  They also suggest that BY adjustments attain similar and overall slightly higher power than e-value procedures, except for large multiverses testing many hypotheses. Hence, the appeal of e-values may lie more in their flexibility to accumulate evidence across hypotheses or across datasets than on FDR control.

The paper is structured as follows. In Section~\ref{sec:framework} we introduce notation and lay out a generalized linear model framework for treatment effect estimation within a multiverse analysis.
Section~\ref{sec:evalues} reviews e-values and their use for multiplicity control  and to aggregate evidence across hypotheses and across datasets. 
Section \ref{sec:simulation} compares several e-value methods  and of BY p-value adjustments  via linear and logistic regression simulations.
Section~\ref{sec:teenager} applies our methodology to a dataset studying the association between teenager mental well-being and technology use.
Section~\ref{sec:discussion} concludes.

\section{Framework}
\label{sec:framework}

We lay out a multiverse framework that is based on generalized linear models and closely resembles that of \cite{semken:2022} but provides FDR control guarantees through e-values.
Consider a setting with $L$ outcomes, $p_x$ treatments and $p_z$ control covariates, with $n$ observations for each. Said observations may be potentially split across $p_g \geq 1$ subpopulations.
 Our setup extends directly to situations where the outcomes are recorded across multiple datasets, each with its own sample size and set of covariates, but we focus on the common $n$ and $(p_x, p_z)$ for simplicity.  
Let $y^{(l)} \in \mathbb{R}^n$ for $l=1,\ldots,L$ be the observed outcomes,
$X$ the $n \times p_x$ matrix and $Z$ the $n \times p_z$ with the treatment and control covariates (respectively),
and $G$ a $n \times p_g$ matrix encoding the subgroup information (if one is not interested in subgroups, then $G$ is omitted).
Let $x_i$, $z_i$ and $g_i$ be the $i^{th}$ row in $X$, $Z$ and $G$ respectively.
Throughout, we denote expectations by $\mathbb E$, probabilities by $\mathbb P$, and by $\mathbb E_P$ and $\mathbb P_P$ expectations and probabilities when data are generated by a distribution $P$.
Further, we denote by $p$ the density (or probability mass) function associated to a distribution $P$, and by $q$ that for a distribution $Q$.

A generalized linear model for outcome $l$ assumes that
\begin{align}
  F( \mathbb E (y_i^{(l)} \mid x_i, z_i, g_i) )=
  \beta_0^{(l)} + z_i^T \eta_z^{(l)} + g_i^T \eta_g^{(l)} + \sum_{j=1}^{p_x} (\beta_j^{(l)} + g_i^T \delta_j^{(l)}) x_{ij},
\label{eq:outcome_equation}
\end{align}
where $F$ is the link function (e.g., the identity for linear regression, or the logit for logistic regression),
$\beta_0^{(l)} \in \mathbb{R}$ is the intercept,
$\beta_j^{(l)} \in \mathbb{R}$ a global coefficient relating treatment $j$ to outcome $l$,
$\delta_j^{(l)}$ captures subpopulation-specific effects,
and $\eta_z^{(l)} \in \mathbb{R}^{p_z}$ and $\eta_g^{(l)} \in \mathbb{R}^{p_g}$ are coefficients associated to control covariates and subgroups respectively.
We denote the whole parameter vector by $\theta^{(l)}= (\{ \beta_j^{(l)} \}_{j=0}^{p_x}, \{ \delta_j^{(l)} \}_{j=1}^{p_x}, \eta_z^{(l)}, \eta_g^{(l)})$.
Importantly, to interpret $\beta_j^{(l)}$ as an average treatment effect, $g_i \in \mathbb{R}^{p_g}$ must encode subpopulation information using a sum-to-zero constraint.
 Specifically, let $\rho_k=n_k/n$ where $n_k$ is the number of observations coming from subpopulation $k$, we define
\begin{align}
  g_{ik}= \begin{cases}
    1 - \rho_k  \mbox{, if observation } i \mbox{ comes from subpopulation } k \\
    - \rho_k \mbox{, otherwise}
    \end{cases}
\nonumber
\end{align}
for $k=1,\ldots,p_g$, which implies that 
\begin{align}
 \sum_{i=1}^n g_{ik}= n_k (1-\rho_k) + (n - n_k) (- \rho_k)= n_k \frac{n - n_k}{n} - (n-n_k) \frac{n_k}{n} = 0.
\nonumber
\end{align}
Intuitively, our coding ensures that the $g_{ik}$'s are orthogonal to the intercept, which as shown below allows interpreting $\beta_j^{(l)}$ as an average measure of association and the terms $g_i^T \delta_j^{(l)}$ as deviations from that average. Specifically,
 we define the effect of treatment $j$ on outcome $l$ for individual $i$ in subpopulation $k$ as
\begin{align}
\beta_{jk}^{(l)}=  \frac{\partial}{\partial x_{ij}} F(\mathbb E (y_i^{(l)} \mid x_i, z_i, g_i)) 
= \beta_j^{(l)} + g_i^T \delta_j^{(l)}
= \beta_j^{(l)} +  (1-\rho_k)  \delta_{jk}^{(l)} - \sum_{m \neq k} \delta_{jm}^{(l)}  \rho_m .
\label{eq:treatment_effect}
\end{align}
The corresponding average treatment effect (ATE) across all individuals is
\begin{align}
\frac{1}{n} \sum_{i=1}^n \frac{\partial}{\partial x_{ij}} F(\mathbb E (y_i^{(l)} \mid x_i, z_i, g_i))
= \beta_j^{(l)} + \frac{1}{n} \left( \sum_{i=1}^n g_i^T \right) \delta_j^{(l)}= \beta_j^{(l)}.
\nonumber
\end{align}

We remark that we use the treatment effect nomenclature without pretension that it necessarily represents a causal effect. Rather, $\beta_j^{(l)}$ measures the conditional association between treatment $j$ and outcome $l$, given control covariates. For example, in our teenager application we find associations between mental well-being and technology use, but we do not claim said associations to be necessarily causal. For $\beta_j^{(l)}$ to identify a causal effect in our framework, two further assumptions are needed (see \cite{hernan:2010}, Chapter 15, for example). Briefly, a first assumption is that the outcome equation \eqref{eq:outcome_equation} specifies correctly the relation between the covariates and the mean of the outcome. The second assumption is that there are no unmeasured confounders.

An alternative definition for the treatment effect to that in \eqref{eq:treatment_effect}, common in the causal inference literature, is the following
\begin{align}
& \frac{\partial}{\partial x_{ij}} \mathbb E (y_i^{(l)} \mid x_i, z_i, g_i)
  = D_i [ \beta_j^{(l)} + g_i^T \delta_j^{(l)} ],
\nonumber \\
&D_i = (F^{-1})' ( \beta_0^{(l)} + z_i^T \eta_z^{(l)} + g_i^T \eta_g^{(l)} + \sum_{j=1}^{p_x} (\beta_j^{(l)} + g_i^T \delta_j^{(l)}) x_{ij} )
\nonumber
\end{align}
where $(F^{-1})'$ denotes the derivative of the inverse link function $F^{-1}$ in \eqref{eq:outcome_equation}.
Here the effect depends on the individual's covariate values $(x_i, z_i)$, and
testing for a treatment effect in subpopulation $g_i$ therefore corresponds to testing $\beta_j^{(l)} + g_i^T \delta_j^{(l)}=0$, as in our previous definition. If one is not interested in subpopulations and drops $g_i$ from the regression equation, then testing $
\beta_j^{(l)}=0$ tests for a non-zero ATE, also as before.


We consider a multiverse analysis where one is interested in estimating the ATEs $\beta_j^{(l)}$ for $l=1,\ldots,L$ and $j=1,\ldots,p_x$,
and performing the corresponding $L p_x$ hypothesis tests
\begin{align}\label{eq:test}
H_{lj}: &\beta_j^{(l)}= 0, \beta_{j'}^{(l)} \neq 0 \mbox{ for } j' \neq j
\nonumber
\\
\overline{H}_l: & \beta_{j'}^{(l)} \neq 0 \mbox{ for } j' \in \{1,\ldots,p_x\}
\end{align}
That is, the null hypothesis $H_{lj}$ is that treatment $j$ has zero effect on outcome $l$, without any restrictions on the remaining treatment effects.
And there is a global alternative hypothesis $\overline{H}_l$, shared across all treatments, that all treatment effects are non-zero.
We denote by $\mathcal{H}_{lj}$ the set of probability distributions such that $H_{lj}$ holds, and by $\overline{\mathcal{H}}_l$ the set where  none of the $H_{lj}$'s hold (that is, $\overline{\mathcal{H}}_l$ is the complement of $\bigcup_{j=1}^{p_x} \mathcal{H}_{lj}$). 
If one is also interested in testing the effect within subpopulations, there are $L p_x p_g$ further tests corresponding to $H_{ljk}: \beta_{jk}^{(l)}=0$ for  $l=1,\ldots,L$, $j=1,\ldots,p_x$ and $k=1,\ldots,p_g$.
To ease notation we drop the subscript $k$ in the exposition below, that is we describe the case where there are no subpopulations, but our framework remains readily applicable when sub-populations are also of interest.

Although not stated explicitly, none of the remaining parameters in $\theta^{(l)}$ are assumed to be zero, neither under the null nor the alternative.
In particular, the control covariates effects $\eta_z$ are always assumed to be non-zero.
This is in contrast to SCA \citep{simonsohn:2020}, 
which as discussed 
computes a median over subsets of control covariates that has no statistical interpretation. 

Importantly, our strategy assumes that the controls are a so-called valid adjustment set \citep{henckel:2022}, that is
that one has excluded colliders and mediators that would lead to identifying a different treatment effect than what one is interested in.
Under such assumption, the main issue is that some of the controls may not be needed because they truly have no effect on the outcome, given the other covariates. Such controls are called instruments in causal inference.
Since including instruments in the outcome equation can lead to substantial variance inflation (for example, see \cite{papaspiliopoulos:2026}), one may attempt to exclude them, as was done in the Bayesian SCA of \cite{semken:2022}.
Specifically,  Bayesian SCA computes a weighted average where weights are given by posterior model probabilities, which as $n \to \infty$ converges to performing inference under a model that includes only the controls that have non-zero effects (under mild conditions).
 While such a strategy leads to power gains as $n \to \infty$, for finite $n$ one can make mistakes in removing controls that are truly associated to the outcome (called confounders in causal inference), and hence one does not get exact type I error control. To avoid these issues, here we include all controls, guaranteeing that no confounders are omitted.

In causal inference, in the rare situation where one knows the causal mechanism linking the treatment and outcome, specified by a directed acyclic graph, \cite{henckel:2022} showed how to obtain the optimal set of control covariates. 
Briefly, one should only include as controls the parents of outcome and the parents of any mediators, excluding mediators and descendants of mediators.
By doing so, one avoids the mentioned variance inflation issues, as well as biases due to colliders and mediators.
These considerations notwithstanding, we reiterate that our framework need not be framed within causal inference. Rather, it is a strategy to study conditional associations (as we do in the teenager application).

Given the potentially large number of multiple tests being conducted, it is important to assess statistical significance in a way such that some measure of false positives is controlled.
Specifically, we consider the FDR \citep{benjamini:1995}. Let $\gamma_{lj}= \mbox{I}(\beta_j^{(l)} \neq 0)$ be the indicator that the null hypothesis $H_{lj}$ does not hold, let $\hat{\gamma}_{lj} = 1$ if we reject $H_{lj}$ and $\hat{\gamma}_{lj}=0$ otherwise. 
Then the number of false positives is $\sum_{l,j} \gamma_{lj} \hat{\gamma}_{lj}$, the number of rejected null hypotheses is $\sum_{l,j} \hat{\gamma}_{lj}$, and $\mbox{FDR}= \mathbb E(\mbox{FDP})$, where 
\begin{align}
 \mbox{FDP}= \begin{cases}
\frac{\sum_{l,j} \gamma_{lj} \hat{\gamma}_{lj}}{\sum_{l,j} \hat{\gamma}_{lj}} \mbox{, if } \sum_{l,j} \hat{\gamma}_{lj} > 0
\nonumber \\
0  \mbox{, otherwise} 
\end{cases}
\nonumber
\end{align}
and the expectation is with respect to the data-generating distribution.
That is, the FDR is the expected false discovery proportion (FDP) and, if no hypotheses are rejected, then FDP is defined to be 0.
If one is interested in testing subpopulation effects $\beta_{jk}^{(l)}$, then the FDR is analogously defined by taking sums over $k$ as well.

\section{e-values}
\label{sec:evalues}

 E-values \citep{vovk:2021,gruenwald:2024,chugg:2026,ramdas:2025} measure evidence against null hypotheses, and allow one to combine evidence across hypotheses and across datasets, while controlling type I errors and the FDR, 
regardless of the dependence structure between the tests.
Formally, an e-value is the realization of a random variable called an e-variable, whose defining property is that its expectation under the null hypothesis is $\leq 1$.

\begin{definition}
Consider a null hypothesis $H_{lj}$ specifying a set of probability distributions $\mathcal{H}_{lj}$.
A test statistic $E_{lj}$ is an e-variable if and only if
$\mathbb E_P (E_{lj}) \leq 1$ for any $P \in \mathcal{H}_{lj}$.
\label{def:evariable}
\end{definition}

Larger values of $E_{lj}$ provide stronger evidence for the alternative hypothesis.
By Markov's inequality, if one rejects $H_{lj}$ when $E_{lj} \geq 1/\alpha$, then the type I error is $\leq \alpha$:
\begin{align}
 \mathbb P_P \left( E_{lj} \geq \frac{1}{\alpha} \right) \leq \alpha \mathbb E_P(E_{lj}) \leq \alpha,
\nonumber
\end{align}
for any $P \in \mathcal{H}_{lj}$.

If $E_{lj}$ and $E_{l'j'}$ are e-variables for the null hypotheses $H_{lj}$ and $H_{l'j'}$, then $(E_{lj} + E_{l'j})/2$ (or any weighted average) is an e-variable under the joint null hypothesis that both $(H_{lj}, H_{l'j'})$ hold. This allows obtaining an aggregate measure of evidence against a set of null hypotheses, for example that a given treatment has no effect on any of the outcomes.
Further, if $E_{lj}$ and $E_{lj}'$ are e-variables obtained in two independent datasets, or in two independent subpopulations, then $E_{lj} E_{lj}'$ is an e-variable. The anytime-validity property of e-values guarantees that one can decide whether to observe $E_{lj}'$ or not after having observed $E_{lj}$: regardless, $E_{lj} E_{lj}'$ remains a valid e-variable, since $\mathbb E [E_{lj} E_{l'j'}] = \mathbb E[ E_{lj} ] \mathbb E [E_{l'j'}] \leq 1$.

In Section~\ref{ssec:define_evariables} we review some strategies to define e-variables,
and in Section~\ref{ssec:fdr_control} methods for FDR control.

\subsection{Defining e-variables}
\label{ssec:define_evariables}

There are many ways to define valid e-variables. While the Markov inequality guarantees type I error control for all, their statistical power may vary greatly. Efforts have been made to define e-variables that accumulate evidence against the null as fast as possible, typically by maximizing $\mathbb E_Q [\ln E_{lj}]$, where $Q \in \overline{\mathcal{H}}_l$ is a distribution within the alternative hypothesis. 
One possible definition of optimality is the so-called numeraire (\cite{ramdas:2025}, Chapter 6), an e-variable $E_{lj}^*$ such that $\mathbb E_Q [ E_{lj}/E_{lj}^*] \leq 1$ under the alternative, where $E_{lj}$ is any other e-variable. 
Another definition by \cite{gruenwald:2024} uses a worst-case optimality criterion called GROW.
When testing a point null hypothesis versus a point null alternative, both definitions return the likelihood-ratio as the optimal e-variable. This is not applicable to our setting, where the null hypothesis $\mathcal{H}_{lj}$ contains infinitely many distributions indexed by $\theta^{(l)}$, and so does the alternative $\overline{\mathcal{H}}_l$.
In such more general settings, obtaining either the numeraire or the GROW-optimal e-variable is related to so-called reverse information projections (RIP), and it is a challenging task.

We review next some practical alternatives applicable to our generalized linear model \eqref{eq:outcome_equation}:
universal mixture e-variables, soft-rank e-variables and p-to-e calibration.
Universal inference \citep{wasserman:2020} 
splits the observed data $y^{(l)}=(y_0^{(l)},y_1^{(l)})$, and $y_0^{(l)}$ is assumed independent from $y_1^{(l)}$. One then obtains the maximum likelihood estimate (MLE) for the density function under the null $\hat{p}_{lj} \in \mathcal{H}_{lj}$ using only $y_0^{(l)}$, that under the alternative $\hat{q}_l$ using only $y_1^{(l)}$, and obtains the split likelihood-ratio e-variable $E_{lj}= \hat{q}_l(y_0^{(l)}) / \hat{p}_{lj}(y_0^{(l)})$ \citep{ramdas:2026}.
Universal mixture e-variables are a variation (\cite{ramdas:2025}, Chapter 5.2) that avoids arbitrary data splits and assuming independence, by considering a hybrid of frequentist and Bayesian hypothesis tests.
Specifically, let $\hat{p}_{lj} \in \mathcal{H}_{lj}$ be the MLE under the null, now using the full data $y^{(l)}$. Define
\begin{align}
E^{um}_{lj}= \frac{q_l(y^{(l)})}{\hat{p}_{lj}(y^{(l)})};
\qquad  q_l(y)= \int p(y \mid \theta^{(l)}) p(\theta^{(l)} \mid \overline{H}_l) d\theta_l
\label{eq:univ_evalue_mixture}
\end{align}
where $q_l(y)$ is the marginal likelihood with respect to a prior density $p(\theta^{(l)} \mid \overline{H}_l)$ that places full support on the alternative 
(i.e., such that the prior probability that $\theta_j^{(l)} \neq 0$ is 1).
$E_{lj}$ is an e-variable for any choice of $p(\theta^{(l)} \mid \overline{H}_l)$.
Since $\hat{p}_{lj}$ is the MLE under $\mathcal{H}_{lj}$, by definition we have that $p(y^{(l)}) \leq \hat{p}_{lj}(y^{(l)})$ for any $P \in \mathcal{H}_{lj}$, therefore
\begin{align}
 \mathbb E_{P} \left( E_{lj} \right)=
\int \frac{q_l(y^{(l)})}{\hat{p}_{lj}(y^{(l)})} p(y^{(l)}) dy^{(l)}
\leq \int \frac{q_l(y^{(l)})}{p(y^{(l)})} p(y^{(l)}) dy^{(l)}= \int q_l(y^{(l)}) dy^{(l)}= 1.
\nonumber
\end{align}
In our framework~\eqref{eq:outcome_equation}, one may consider Zellner's unit information prior \citep{zellner:1986}, a classical default with tight connections to the Bayesian information criterion \citep{schwarz:1978}.
Specifically, let $W=(X, Z)$ be the design matrix associated to the regression in~\eqref{eq:outcome_equation} (if subpopulations are of interest, then subgroups $G$ and interaction terms $GX$ are also included in $W$).
Zellner's unit information prior is
\begin{align}
p(\theta^{(l)}  \mid \sigma_l^2, \overline{H}_l)  = N(\theta^{(l)}; 0, n \sigma_l^2 (W^T W)^{-1}),
\label{eq:uip}
\end{align}
where $\sigma_l^2 > 0$ is the GLM dispersion parameter associated to $y^{(l)}$.
 For some models such as logistic or Poisson regression $\sigma_l=1$ is fixed, while in others such as linear regression $\sigma_l$ is unknown and sets a prior on it, for example a minimally informative $\sigma_l^2 \sim \mbox{InvGamma}(0.01/2, 0.01/2)$. The interpretation that \eqref{eq:uip} contains as much Fisher information as a sample of size 1 only holds exactly in linear regression. For other GLMs the Fisher information is $W^T D(\theta_0) W$, where $D(\theta_0)$ is a diagonal matrix that depends on the data-generating $\theta_0$. Since $\theta_0$ is unknown, we use $W^T W$ as a simple proxy. Another simple alternative is to use $W^T D(0) W$ \citep{sabanes:2011}. 
Our choice assumes that $W^T W$ is invertible, 
otherwise an alternative is to replace $W^TW$ by $\mbox{diag}(W^T W)$. Expression~\eqref{eq:univ_evalue_mixture} requires obtaining an MLE for the $L p_x$ alternative hypotheses, which can be done in parallel using any suitable optimization algorithm.
It also requires obtaining $L$ marginal likelihoods $q_l(y)$, one for each outcome.
In the linear regression case where $y^{(l)} \sim N(W \theta^{(l)}, \sigma_l^2 I)$, $q_l$ has a closed-form.
Specifically, if one sets an inverse Gamma prior $\sigma_l^2 \sim \mbox{IG}(a/2, b/2 )$, then
\begin{align}
 \frac{q_l(y^{(l)})}{\hat{p}_{lj}(y^{(l)})}= 
\frac{ b^{a/2}  \Gamma(\frac{n+a}{2})}{(n + 1)^{d/2} \Gamma(a/2)}
\frac{(2 e/\hat{\sigma}_{lj}^2)^{\frac{n}{2}}}{\left(  b  + \| y^{(l)} \|^2 - \frac{n}{n+1} (\hat{\theta}^{(l)})^T W^T W \hat{\theta}^{(l)} \right)^{\frac{n + a}{2}}}
\nonumber
\end{align}
where $\| \cdot \|^2$ is the squared Euclidean norm, $d=\mbox{dim}(\theta^{(l)})$, $\hat{\theta}^{(l)}= (W^T W)^{-1} W^T y^{(l)}$ is the least-squares estimator and $\hat{\sigma}_{lj}^2$ the MLE for the error variance under the null $H_{lj}$.
Outside of linear regression numerical integration methods must be used and, as long as the number of outcomes $L$ is not too large, this can be done quickly (and in parallel).
One may also use Laplace approximations to $q_l$, which requires a single optimization for each outcome, and is very accurate as $n$ grows (\cite{kass:1990}, \cite{rossell:2021b} Lemma S3).

A second strategy is to use soft-rank e-variables \citep{koning:2023}. 
These are typically based on permutations akin to those used to obtain permutation-based p-values, therefore they are more computationally-intensive but as a positive aspect they are more reliable when our statistical model in \eqref{eq:outcome_equation} is misspecified (e.g., truly non-linear covariate effects, heteroskedastic errors,  or a misspecified link function ).
Let $T_0 \geq 0$ be a test statistic for the null hypothesis of interest, and let $T_1,\ldots,T_B$ be the realization of $B$ test statistics that are exchangeable with $T_0$ under the null. The soft-rank e-variable is 
\begin{align}
  E^{sr} \,=\,\frac{B T_0}{\sum_{b=1}^{B}T_b}.
\end{align}
In GLMs, 
one can use the likelihood ratio $T_0=\hat{q}_l(y^{(l)}) / \hat{p}_{lj}(y^{(l)})$
associated to \eqref{eq:test},
where recall that $\hat{p}_{lj}$ is the MLE for the density under the null $H_{lj}$, and $\hat{q}_l$ that under the alternative. 
The exchangeable test statistics $T_1,\ldots,T_B$ can be obtained using permutations. 
Let $y^{(l,b)}$ be a permuted version of $y^{(l)}$ such that the null hypothesis $H_{lj}$ holds, and define
$T_b=\hat{q}^{b}_l(y^{(l,b)}) / \hat{p}^b_{lj}(y^{(l,b)})$
where $\hat{p}^b_{lj}$ and $\hat{q}^{b}$ are the MLEs under the null and alternative hypotheses respectively, computed on the permuted dataset $b$. Although we denote permutations by $y^{(l,b)}$ for simplicity, permutations typically affect covariate values rather than the outcome. For example, in our simulations, permutations are obtained by first regressing the treatment of interest on all other covariates, permuting the residuals from such regression, and defining a synthetic treatment given by the expected value given other covariates plus a permuted residual. In such a permutation, the outcome is conditionally independent from the treatment given other covariates, hence $T_0$ is exchangeable with $T_1,\ldots,T_B$ under $H_{lj}$.

The third and final option that we discuss here to obtain an e-variable is to use so-called p-to-e calibrators \cite{ramdas:2025}, which convert p-values into e-values. Let $\pi_{lj}$ be a p-value for the null hypothesis $H_{lj}$ in \eqref{eq:test}, two popular calibrators are
\begin{equation}
  E^{c1}(\pi_{lj})=\int_0^1 \kappa \pi_{lj}^{\kappa-1} \mathrm{~d} \kappa=\frac{1-\pi_{lj}+\pi_{lj} \ln \pi_{lj}}{\pi_{lj}(\ln \pi_{lj})^2};
\qquad   E^{c2}(\pi_{lj})=\pi_{lj}^{-1/2} -1.
\label{eq:calibrator1}
\end{equation}
We refer to $E^{c1}$ as the mixture calibrator,
as it uses a mixture over $\kappa\in(0,1)$ of the calibrator by \cite{shafer:2011}, 
and to $E^{c2}$ as the inverse root calibrator.
The family of p-to-e calibrators is much richer, with no single calibrator uniformly outperforming all others. See \cite{ramdas:2025} (Chapter 2) for a review. We take $\pi_{lj}$ to be the likelihood ratio test p-value for~\eqref{eq:test}. 
A theoretical caveat is that, outside of the Gaussian family, most implementations of $\pi_{lj}$ rely on an asymptotic chi-square approximation. 
Since such p-value is not exact for finite $n$, neither is the calibrated e-value.
However, in our simulations \eqref{eq:calibrator1} controlled the type I error adequately, for moderately large $n$.

\subsection{FDR control with e-variables}
\label{ssec:fdr_control}

We first review strategies to control the FDR with e-variables, and next compare them with BY p-value adjustment \citep{benjamini:2001}, which also controls FDR under arbitrary dependences between the tests.

For e-variables, a simple strategy to control the FDR below any user-specified $\alpha$ is the e-BH procedure of \cite{wang_ruodu:2022}. An alternative that strictly dominates e-BH in terms of power is the closed e-BH procedure (\cite{ramdas:2025}, Chapter 9.6), 
albeit the method and computations are slightly more involved.
The e-BH procedure proceeds as follows.
Let $e_{lj}$ be the e-value, that is the observed value of the e-variable $E_{lj}$, for $l=1,\ldots,L$ and $j=1,\ldots,p_x$, and let $K= L p_x$ be the number of hypothesis tests.
Let $e_{[k]}$ be the $k^{th}$ order statistic of $e_{lj}$ across all $(l,j)$, sorted from largest to smallest. The e-BH procedure at level $\alpha$ rejects all hypotheses with the $k^*$ largest e-values, where
\begin{align}
 k^*= \max \left\{ k \in \{1, \ldots, L p_x \} : \frac{k e_{[k]}}{K} \geq \frac{1}{\alpha}  \right\},
\nonumber
\end{align}
with the convention that $\max \emptyset = 0$.
That is, the $k^{th}$ test effectively uses $k e_{[k]} / K$ as its e-value, and if we reject test $k$ then we also reject all the previous tests.
\cite{wang_ruodu:2022} proved that 
$\mbox{FDR} \leq \alpha K_0 / K \leq \alpha$, where $K_0$ is the number of truly null hypotheses.


We now describe the closed e-BH procedure.
Let $R_k$ be the set of indices corresponding to the $k$ largest e-values, closed e-BH rejects all null hypotheses $k \leq \bar{k}$, where
$\bar k= \max \{k \geq 0: R_k \in C_\alpha \}$ and
\begin{align}
 C_\alpha = \left\{ R: \frac{1}{|A|} \sum_{k \in A} E_k \geq \frac{\mbox{FDP}_A(R)}{\alpha}  \mbox{, for all } A \subseteq \{ 1,\ldots,K \}  \right\}
\nonumber
\end{align}
where $\mbox{FDP}_A(R)= |A \cap R|/\max\{|R|, 1\}$ is the false discovery proportion associated to rejecting the hypotheses in $R$, if $A$ were the subset of truly null hypotheses.
In words, $R_{\bar k}$ is such that the average of the e-values over any other subset $A$ exceeds $\mbox{FDP}_A(R)/\alpha$.
Closed e-BH 
rejects at least as many hypothesis as e-BH.
However, finding $R_{\bar k}$ requires order $K^3$ operations, whereas e-BH only requires $K$ operations.

We now compare these procedures with BY, which uses p-values rather than e-variables.
If one is solely interested in FDR control, whether e-variables are preferable to BY or not depends on their statistical power.
Current evidence suggests that no approach uniformly dominates the other.
\cite{wang_ruodu:2022} showed that e-BH can have higher power than BY in certain settings and lower on others, depending on how the e-variables are obtained and on their distribution.
To gain some intuition, let $\pi_{[k]}$ be the $k^{th}$ ordered p-value.
BY rejects the largest $k$ such that
\begin{align}
\pi_{[k]} \leq \frac{\alpha k}{K c(K)}
\nonumber
\end{align}
where $c(K)= \sum_{i=1}^K 1/i \approx \ln(K) + 0.577$ is the $K^{th}$ harmonic number. 
In contrast, e-BH rejects the largest $k$ such that
$
(1 / e_{[k]}) \leq \alpha k/ K.
$

There are therefore two issues driving these two procedures.
First, as pointed out by \cite{wang_ruodu:2022}, e-BH avoids the $\ln(K)$ term in the denominator, hence the threshold scales better as $K$ grows. 
Second, for any fixed $K$ the relative powers depend on the magnitude of $1/e_{[k]}$ versus that of $\pi_{[k]}$. For example, for the mixture calibrator e-variable (which had a fairly competitive power in our experiments), we have
\begin{align}
\lim_{\pi_{[k]} \to 0} e_{[k]}= \lim_{\pi_{[k]} \to 0} \frac{1 - \pi_{[k]} + \pi_{[k]} \ln(\pi_{[k]})}{\pi_{[k]} (\ln \pi_{[k]})^2}
= \frac{1}{\pi_{[k]} (\ln \pi_{[k]})^2}.
\nonumber
\end{align}
Hence and roughly speaking, for small $\pi_{[k]}$, applying e-BH to the mixture calibrator seeks the largest $k$ such that
\begin{align}
 \pi_{[k]} (\ln \pi_{[k]})^2 \leq \frac{\alpha k}{K}.
\nonumber
\end{align}
Therefore, if $(\ln \pi_{[k]})^2 > \ln(K) + 0.577$ then one expects BY to have higher power than e-BH, and otherwise to have lower power. Note that the rejected hypotheses are driven by the largest $k$ satisfying these thresholds, hence power is not driven by the most extreme e-variables / p-values.
Recall also that closed e-BH has higher power than e-BH, though in our experiments both delivered almost identical results.

As a word of caution, in sequential settings where one decides to observe more data based on previous e-values, a repeated application of FDR procedures does not guarantee FDR control. For example, applying e-BH only once after all data has been obtained controls the FDR, since for each individual test one has an e-variable, but applying e-BH to a first dataset and then applying e-BH again after observing a second dataset does not control the overall FDR across both applications of e-BH.

\section{Simulation study}
\label{sec:simulation}

Since there are several strategies to define e-variables (Section \ref{ssec:define_evariables}), we first compare in Section \ref{ssec:simstudy_evalues} their statistical powers via simulations in single-outcome linear and logistic regressions, featuring multiple treatments and controls.
In Section \ref{ssec:simulation_multiverse} we extend the simulations to a multiverse setting with several outcomes, focusing on the two most powerful e-values found in Section \ref{ssec:simstudy_evalues}, and compare them to BY p-value adjustment.

\subsection{Comparing e-variables}\label{ssec:simstudy_evalues}

\begin{figure}[ht]
\begin{tabular}{cc}
           $\beta_i^*=0$& $\beta_i^*=0.15$ \\
         \includegraphics{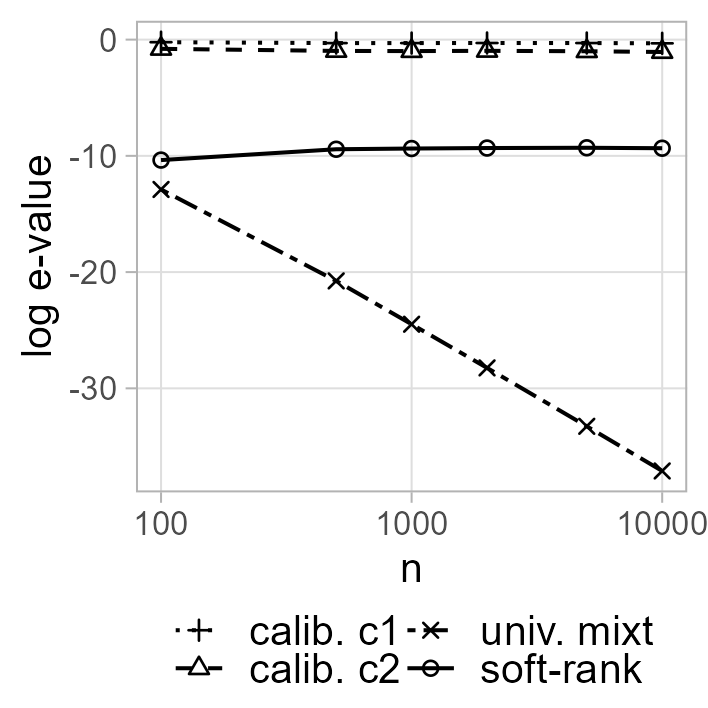}  & 
         \includegraphics{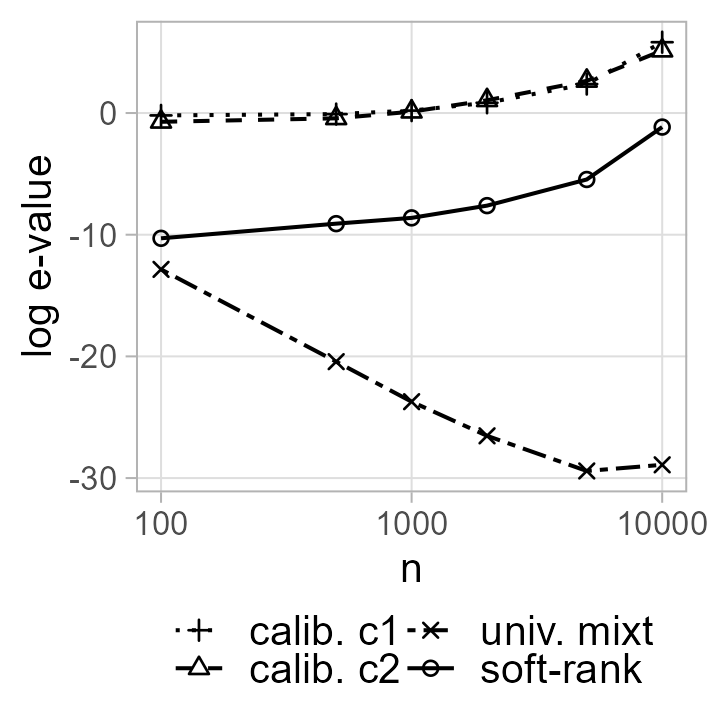} \\
         $\beta_i^*=0.3$& $\beta_i^*=0.5$ \\
         \includegraphics{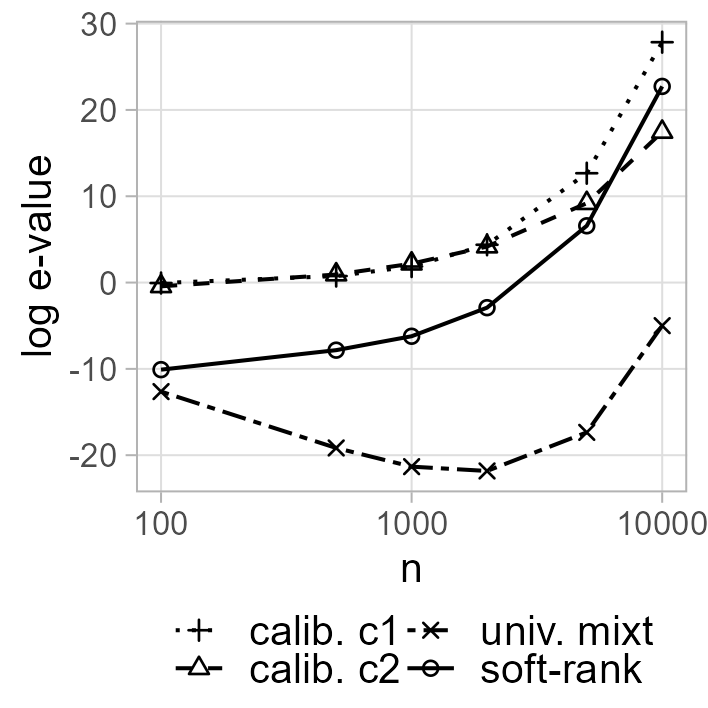}  & 
         \includegraphics{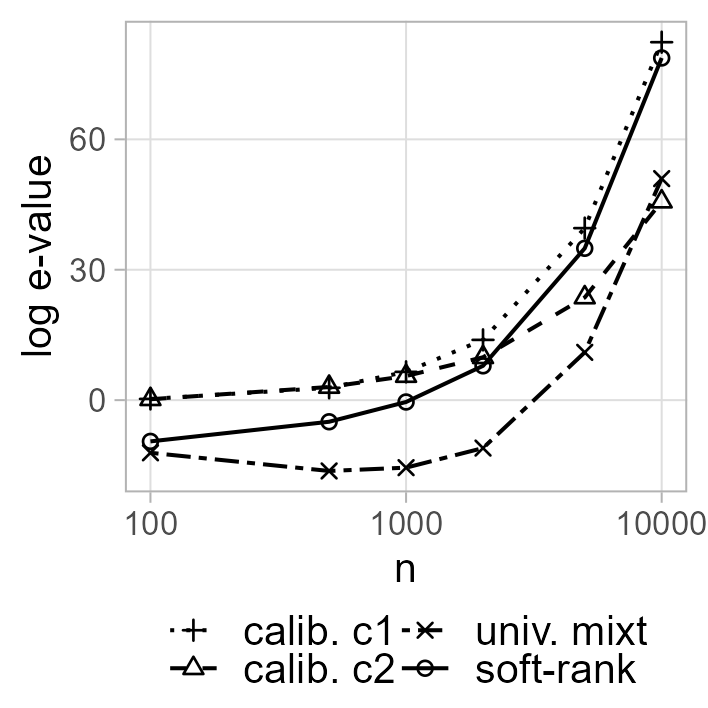} \\
\end{tabular}
\caption{Average log e-values in logistic regression for $\beta_i^*=0$ (top left), $\beta_i^*=0.15$ (top right), $\beta_i^*=0.3$ (bottom left) and $\beta_i^*=0.5$ (bottom right) for universal mixture and soft-rank e-values, mixture (calib.c1) and inverse root p-to-e calibrators (calib.c2)}
\label{fig:simlogreg}
\end{figure}

\begin{table}[ht]
\caption{Proportion of type I errors { out of 1,000 simulations} for $E\in\{E^{um},E^{sr},E^{c1},E^{c2}\}$ and significance levels $\alpha \in \{0.01,0.05\}$ in logistic regression}\label{tabtype1logreg}%
\begin{tabular}{ccc|cc|cc|cc}
\hline
n    & \multicolumn{2}{c|}{$E^{um}$} & \multicolumn{2}{c|}{$E^{sr}$}& \multicolumn{2}{c|}{$E^{c1}$} & \multicolumn{2}{c}{$E^{c2}$}\\ \hline
     & $\alpha=1\%$ & $\alpha=5\%$ & $\alpha=1\%$ & $\alpha=5\%$ & $\alpha=1\%$ & $\alpha=5\%$ & $\alpha=1\%$ & $\alpha=5\%$ \\
100  & 0                          & 0                          & 0                          & 0                          & 0.001                       & 0.003                       & 0.001                      & 0.006                      \\
500  & 0                          & 0                          & 0                          & 0                          & 0                          & 0.001                          & 0                          & 0.002                          \\
1000 & 0                          & 0                          & 0                          & 0                          & 0                          & 0.001                          & 0                          & 0.002                           \\
2000 & 0                          & 0                          & 0                          & 0                          & 0                          & 0.003                         & 0                          & 0.003                          \\ \hline
\end{tabular}
\end{table}

We assess the type I error and power of the four e-values from Section \ref{ssec:define_evariables}: universal mixture and soft-rank e-values and the mixture and inverse root calibrators in \eqref{eq:calibrator1}. We run a simulation study in linear and in logistic regression. We considered sample sizes $n \in [100, 10000]$ and 10 covariates generated from a zero-mean multivariate Gaussian with pairwise correlations $0.5$, and scaled to unit sample variance. For each sample size, we simulate  1,000  datasets with data-generating regression coefficients  $\beta^*= (0, 0.05, 0.1, 0.15, 0.2, 0.25, 0.3, 0.5, 0.75, 1)$.


\begin{figure}[ht]
\begin{tabular}{cc}
         $\beta_i^*=0$& $\beta_i^*=0.15$ \\
         \includegraphics{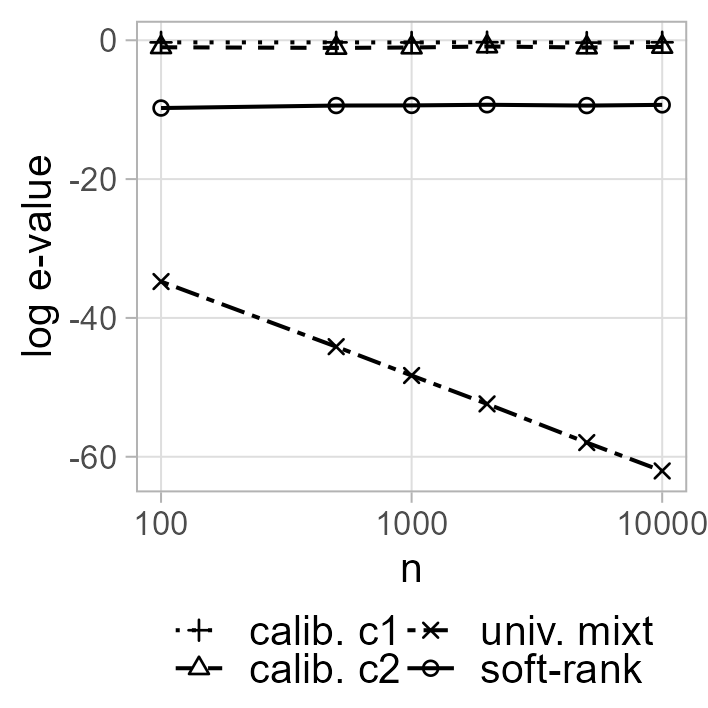}  & 
         \includegraphics{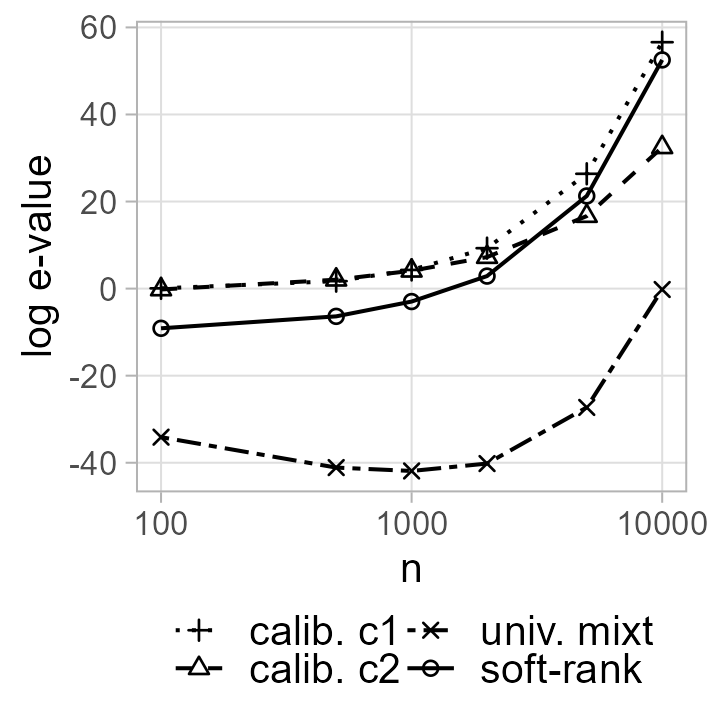} \\
         $\beta_i^*=0.3$& $\beta_i^*=0.5$ \\
         \includegraphics{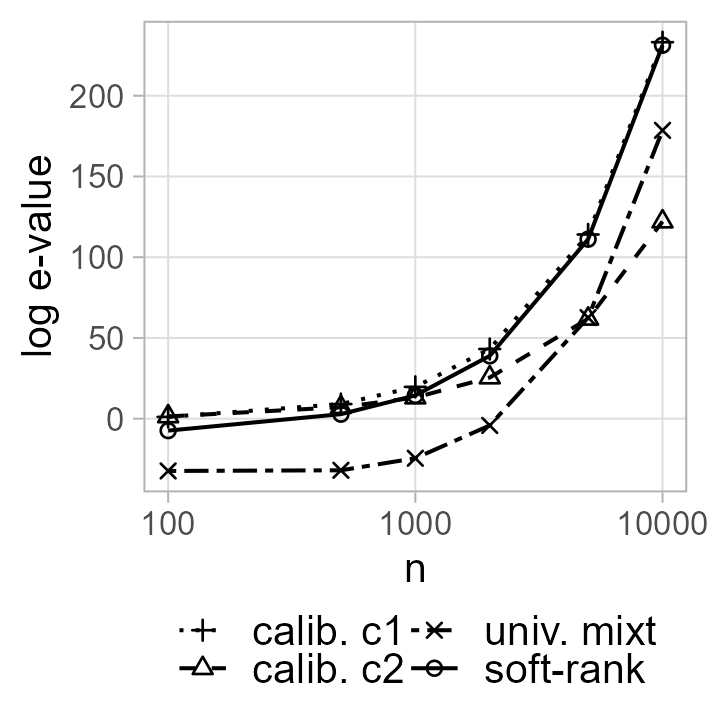}  & 
         \includegraphics{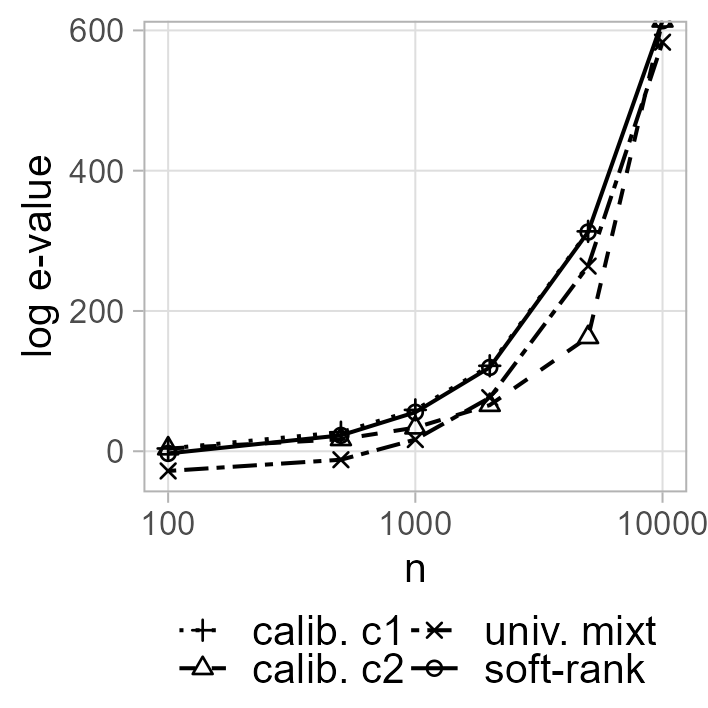} \\
\end{tabular}
\caption{Average log e-values in linear regression for $\beta_i^*=0$ (top left), $\beta_i^*=0.15$ (top right), $\beta_i^*=0.3$ (bottom left) and $\beta_i^*=0.5$ (bottom right) for universal mixture and soft-rank e-values, mixture (calib.c1) and inverse root p-to-e calibrators (calib.c2)}
\label{fig:simlinreg}
\end{figure}

Table~\ref{tabtype1logreg} reports the proportion of type I errors over the  1,000  repetitions 
in logistic regression. All e-variables control the type I error below the significance level $\alpha$, in fact they  were very conservative in that they made very few type I errors. Intuitively, their anytime-validity implies that one can peek at the data at any moment in time, requiring one to be conservative. In Section \ref{ssec:simulation_multiverse} we assess whether this conservativeness significantly reduces statistical power relative to BY p-value adjustment, finding that the effect is fairly minor. 
Figure~\ref{fig:simlogreg} plots the average log e-values 
in logistic regression. Recall that larger e-values indicate stronger evidence against the null. For data-generating $\beta_i^* \neq 0$ and small $n$, the calibrated e-values $E^{c1}$ and $E^{c2}$ clearly dominate the universal mixture and soft-rank e-variables $E^{um}$ and $E^{sr}$, accumulating stronger evidence for the alternative. For large $n$, the calibrated e-variable $E^{c2}$ became less competitive whereas the soft-rank e-variable $E^{sr}$ became more competitive, nearly matching the mixture calibrated e-variable $E^{c1}$ in some scenarios. 
Figure~\ref{fig:simlinreg} reports analogous results for linear regression. 
Here there were no type I errors. 

In summary, 
we recommend the mixture calibrator $E^{c1}$ 
as it accrues stronger evidence for truly alternative hypotheses. 
We remark that such e-values are only calibrated when the assumed model is correct, otherwise permutation-based soft-rank e-variables may be preferable for being more robust to model misspecification.

\begin{table}[ht]
  \caption{ Median computing time in seconds over 1,000 repetitions to obtain the e-values corresponding to the 10 coefficients $\beta_i$, $i=1,\ldots,10$ in linear and logistic regression }
  \label{tabtimereg}%
\begin{tabular}{c|ccc|ccc}
\hline
n    & \multicolumn{3}{c|}{linear} & \multicolumn{3}{c}{logistic} \\ \hline
     & $E^{um}$ & $E^{sr}$ & $E^{c1}$, $E^{c2}$ & $E^{um}$ & $E^{sr}$ & $E^{c1}$, $E^{c2}$  \\
100  & 0.004 & 2.399 & 0.002 & 0.046 & 12.036 & 0.029\\
500  & 0.004 & 2.667 & 0.003 & 0.062 & 20.167 & 0.038\\ 
1,000 & 0.005 & 2.999 & 0.003 & 0.083 & 31.439 & 0.047\\
2,000 & 0.007 & 3.916 & 0.004 & 0.112 & 50.225 & 0.067\\
5,000 & 0.011 & 6.354 & 0.007 & 0.203 & 111.91 & 0.129\\
10,000 & 0.016 & 9.437 & 0.009 & 0.372 & 190.630 & 0.241\\ \hline
\end{tabular}
\end{table}

Table~\ref{tabtimereg} reports the median computing time 
to obtain the 10 e-values corresponding to $\beta_1,\ldots,\beta_{10}$. 
We report a single time for $E^{c1}$ and $E^{c2}$ because they are computed from the same p-value. Computing times are consistently below 1 second for the universal mixture $E^{um}$ and the calibrated e-variables $E^{c1}$ and $E^{c2}$, 
and much longer for the soft rank e-variable $E^{sr}$ ($>$ 3 minutes in logistic regression using 10,000 permutations).

\subsection{Multiverse analysis simulation}
\label{ssec:simulation_multiverse}

\begin{figure}[ht]
\begin{tabular}{c}
$L=5$ outcomes \\
\includegraphics[width=0.8\textwidth, height=0.5\textwidth]{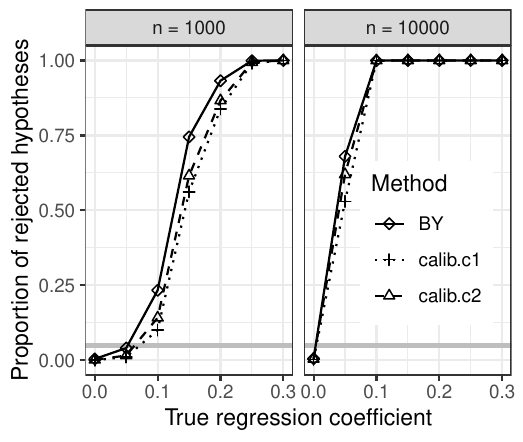}  \\ 
$L=20$ outcomes \\
\includegraphics[width=0.8\textwidth, height=0.5\textwidth]{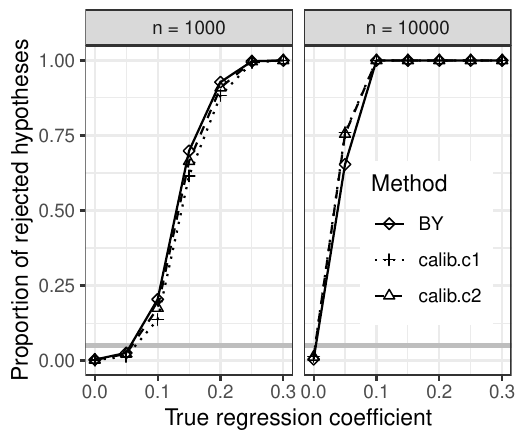}  \\ 
\end{tabular}
\caption{Proportion of rejected hypothesis vs. data-generating $\beta_i^*$ in the multiverse simulations with 5 and 20 outcomes. The thick horizontal line indicates 0.05
}
\label{fig:sim_multiverse_power}
\end{figure}

\begin{figure}[ht]
\begin{tabular}{c}
\includegraphics[width=0.8\textwidth, height=0.5\textwidth]{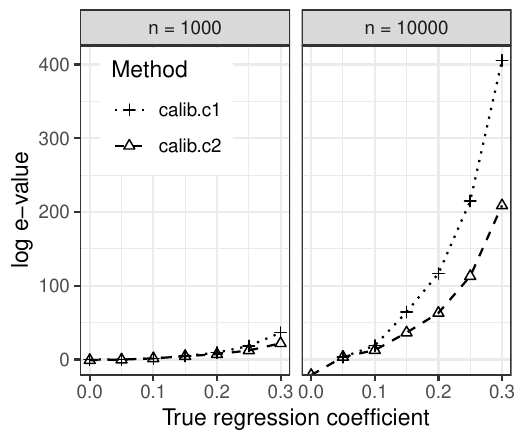} \\
\end{tabular}
\caption{Average log e-values for mixture (calib.c1) and inverse root p-to-e calibrators (calib.c2)
in the multiverse simulation (5 outcomes)
}
\label{fig:sim_multiverse_logevid}
\end{figure}

We consider a setting that mimics the teenager application in Section \ref{sec:teenager}, with $p_x=5$ treatments and $p_z=14$ controls, whose values were generated drawing with replacement observations from the teenager dataset.
We considered $n \in \{1000, 10000\}$,
a moderately-sized multiverse featuring $L=5$ outcomes and a larger multiverse with $L=20$, leading to a total of 25 and 100 hypotheses being tested (respectively).

We simulated the outcomes from linear regression models with independent standard Gaussian errors. The control covariate coefficients were $\eta_z^{(l)}= (1, 0.9, 0.8, \ldots, 0.1, 0, 0, 0, 0)$ for all outcomes. In the multiverse with $L=5$ outcomes, we set the treatment effects to
$\beta^{(1)}= (0, 0, 0, 0, 0)$, 
$\beta^{(2)}= (0, 0, 0.05, 0.1, 0.15)$,
$\beta^{(3)} = (0, 0, 0.1, 0.15, 0.2)$,
$\beta^{(4)} = (0, 0, 0.15, 0.2, 0.25)$,
$\beta^{(5)} = (0, 0, 0.2, 0.25, 0.3)$.
That is, all treatment effects are 0 for outcome 1, and for other outcomes three effects are non-zero.
In the multiverse with $L=20$ outcomes, we used the same $(\beta^{(1)},\ldots,\beta^{(5)})$ for outcomes 1-5, and again the same coefficients for outcomes 6-10, 11-15, and 16-20. 

We considered the mixture $E^{c1}$ and inverse root $E^{c2}$ calibrated e-variables, which exhibited highest power in Section \ref{ssec:simstudy_evalues}, combined with the closed e-BH procedure. We also considered BY-adjusted p-values.

For each $n$ and $L$ combination we simulated 1,000 independent datasets, and we report averaged results.
Both $E^{c1}$ and $E^{c2}$ had a very similar statistical power (Figure \ref{fig:sim_multiverse_power}).
The power of BY-adjusted p-values was also similar, albeit slightly higher in the universe with $L=5$ outcomes and $n=1,000$, and slightly lower for $L=20$ and $n=10,000$.
This agrees with our discussion in Section \ref{ssec:fdr_control}, where e-values may be slightly more powerful than BY when testing many hypotheses.
Figure \ref{fig:sim_multiverse_logevid} shows that $\mathbb E [\ln E^{c1}]$ was higher than $\mathbb E[ \ln E^{c2}]$ (similar to Section \ref{ssec:simstudy_evalues}), the difference was mostly in situations where both were large, and therefore this had a negligible effect on the statistical power at a 0.05 FDR cutoff.

Purely in terms of FDR control, in these simulations there are essentially no benefits in using e-variables over BY-adjusted p-values.
In such settings, the main interest of e-variables may be in easily combining evidence across hypotheses, and across independent datasets, as illustrated in Section \ref{ssec:teenager_combining}.

\section{Teenager mental well-being and technology use}
\label{sec:teenager}

\begin{table}[ht]
\begin{tabular}{@{}lrr@{}}
\toprule
  & Outcome reported & Complete observations \\
\midrule
 Depressed (adolescent)                  & 11,200 & 8,351 \\ 
 Low self-esteem (adolescent)            & 11,165 & 8,311 \\ 
 High total difficulties (parent)        & 11,471 & 8,238 \\ 
 High emotional problems (parent)        & 11,486 & 8,240 \\ 
 High conduct problems (parent)          & 11,488 & 8,243 \\ 
 High hyperactivity/inattention (parent) & 11,481 & 8,243 \\ 
 High peer problems (parent)             & 11,491 & 8,244 \\ 
 Low pro-sociality (parent)              & 11,489 & 8,243 \\
\botrule
\end{tabular}
\caption{ Number of observations and of complete observations (no missings in the treatments nor in the controls) for each outcome in the MCS dataset }
\label{tab:missings}
\end{table}

\begin{figure}[ht]
\begin{tabular}{ccc}
         Depressed & Low self-esteem\\
         \includegraphics{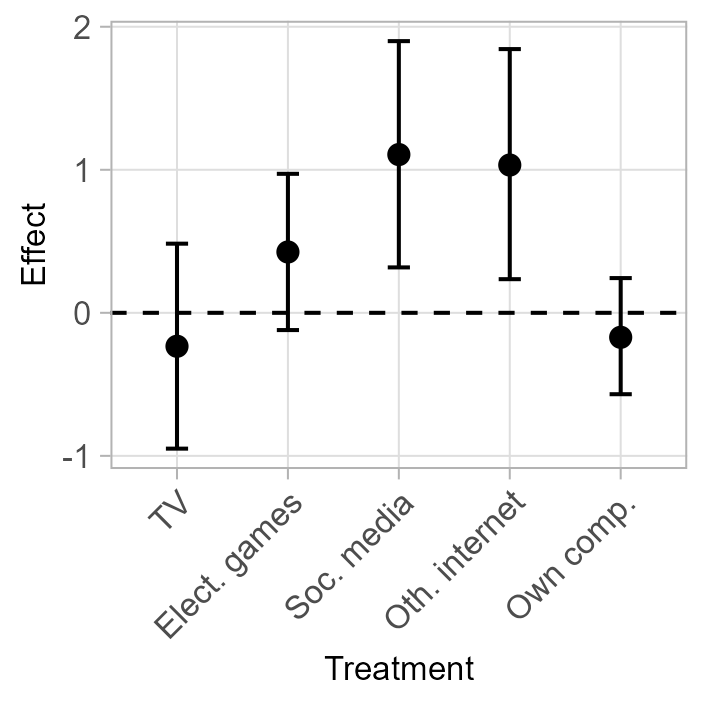}  & 
         \includegraphics{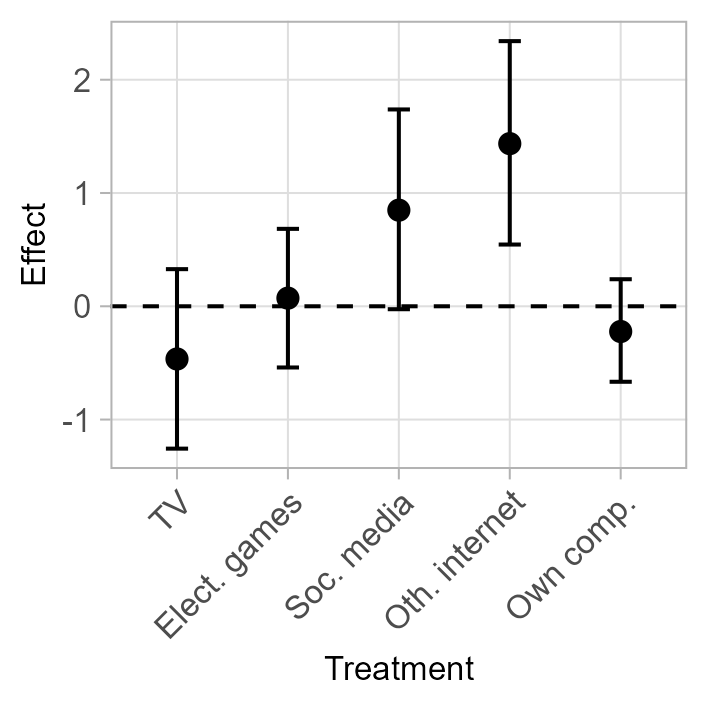} \\
         High total difficulties & High emotional problems\\
         \includegraphics{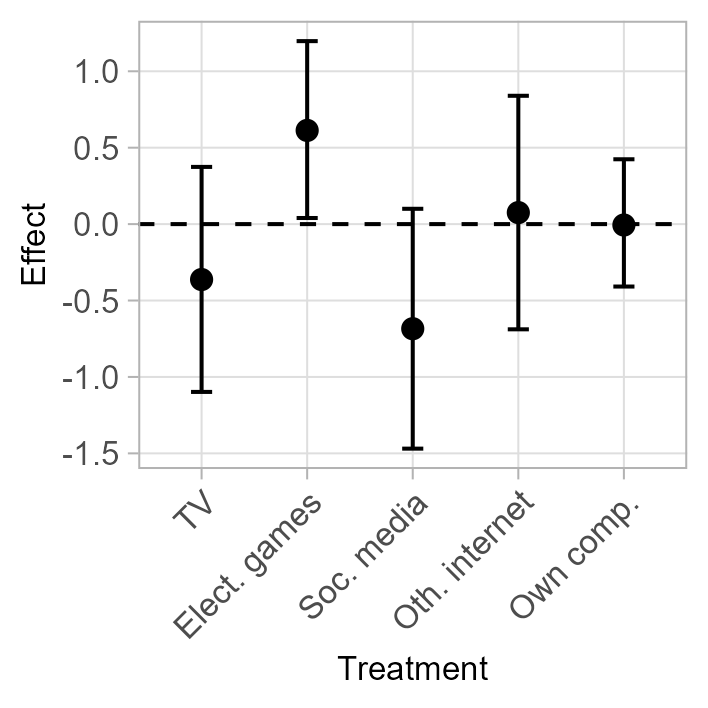}  & 
         \includegraphics{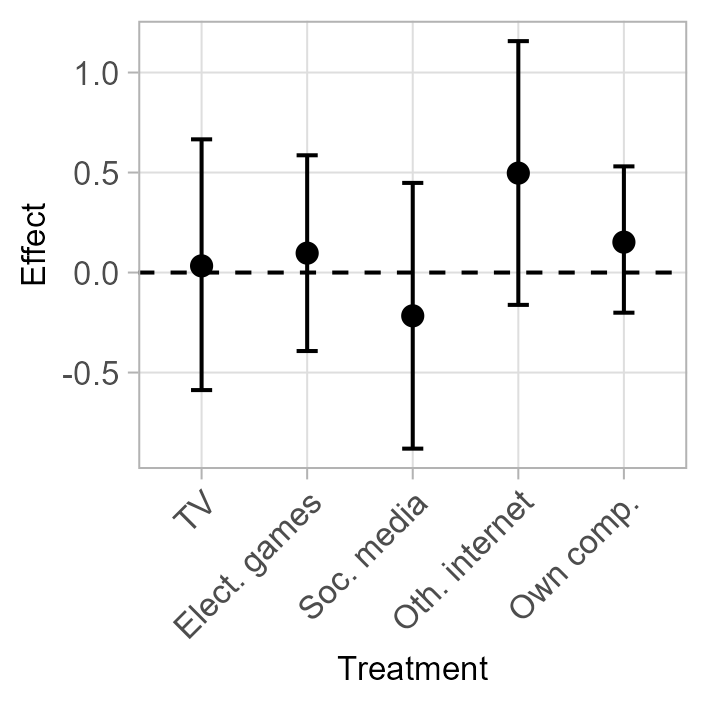} \\
         High conduct problems& High hyperactivity/inattention\\
         \includegraphics{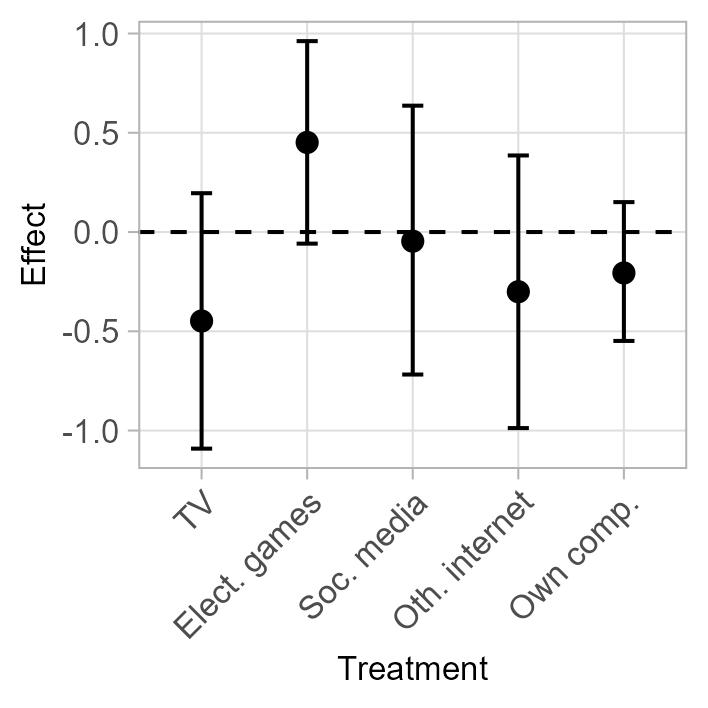}  & 
         \includegraphics{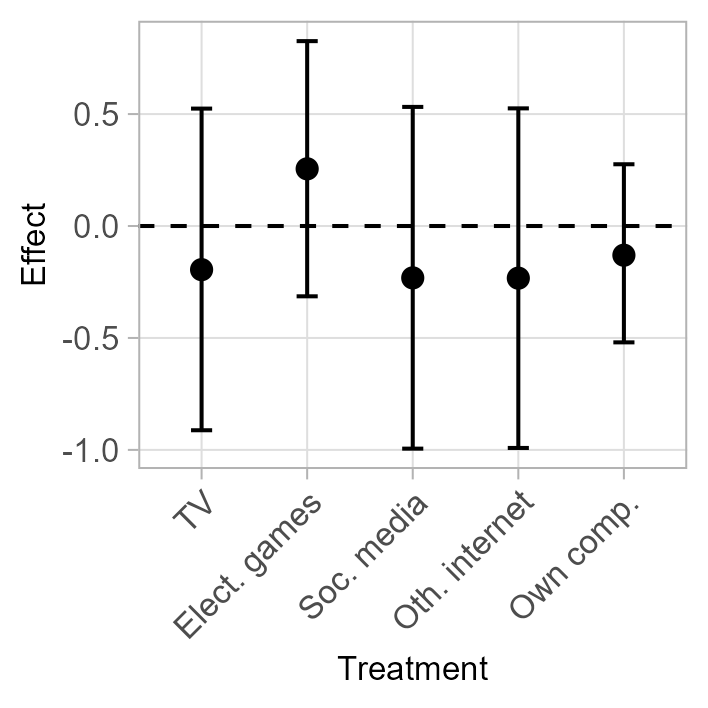} \\
         High peer problems & Low pro-sociality\\
         \includegraphics{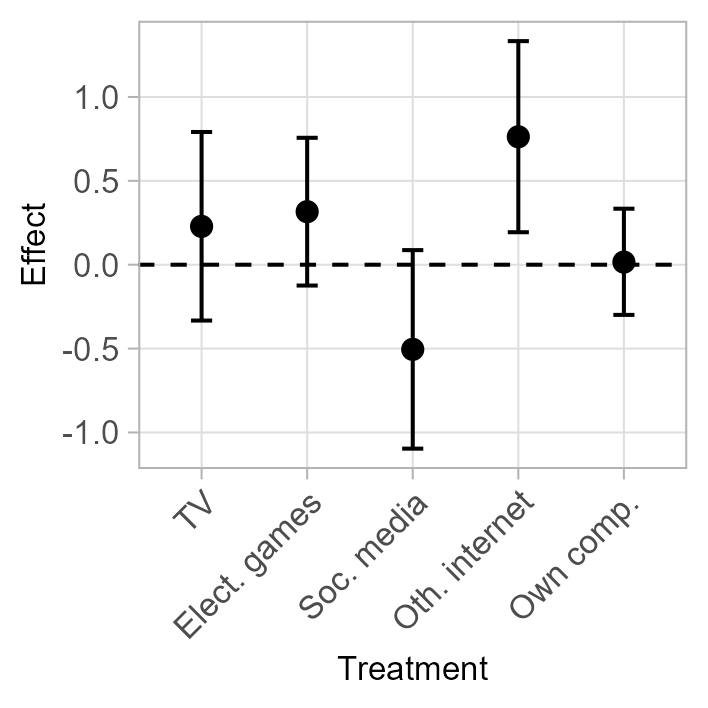}  & 
         \includegraphics{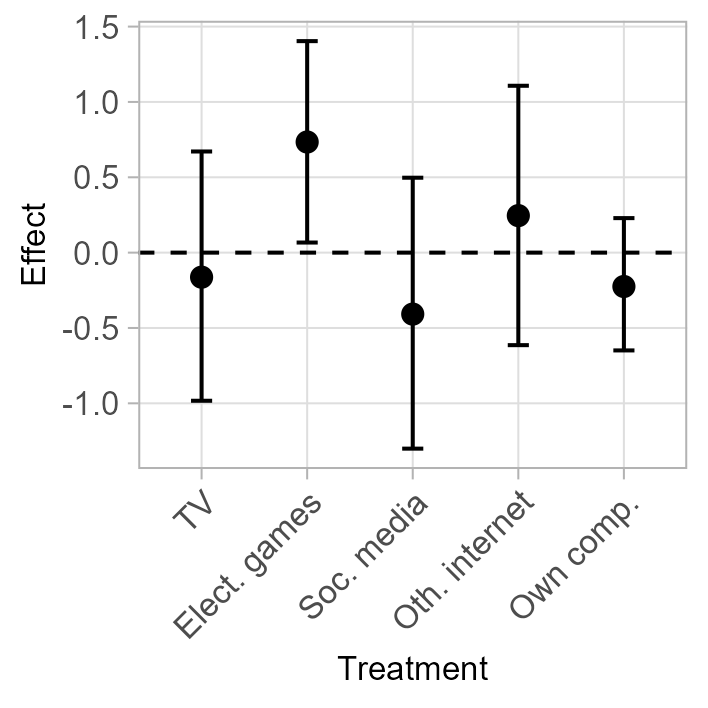}
        \end{tabular}
\caption{Estimated treatment effects by outcome under consideration and their corresponding 95\% e-confidence interval based on $E^{c1}$.}
\label{fig:exp.mcs}
\end{figure}

\begin{table}[ht]
\caption{MCS Data. Rejected hypothesis by the closed e-BH procedure using mixture calibrated e-variables $E^{c1}$ (top 4 rows) or by BY p-value adjustment (top 7 rows) with FDR level 0.05, estimated odds ratio and 95\% e-confidence intervals}\label{tab:mcs_evalues}%
\begin{tabular}{@{}llccccc@{}}
\toprule

Outcome & Treatment & Odds ratio & 95\% conf. int. & e-value  &  BY p-value\\
\midrule
Low self-esteem     &  internet        & 4.20 &(1.72,10.39) & 37,790&  1.75 $\times 10^{-5}$ \\
Depressive symptoms & social media     & 3.02 &(1.37,6.68)  & 1,851 &  2.92 $\times 10^{-4}$ \\
High peer problems  & internet         & 2.14 &(1.21,3.79)  & 757   &  5.68 $\times 10^{-4}$ \\ 
Depressive symptoms &  internet        & 2.81 &(1.27,6.32)  & 396   &  9.38 $\times 10^{-4}$ \\
Low pro-sociality   & electronic games & 2.08 & (1.07,4.07) &  52.6 &  0.010  \\
Total difficulties  & electronic games & 1.85 & (1.04,3.31) &  36.9 &  0.013 \\
Low self-esteem     & social media     & 2.34 & (0.97,5.68) &  15.7 &  0.036 \\
\botrule
\end{tabular}
\end{table}

We apply our framework to study the association between technology and teenager mental well-being in the 2015 survey of the Millennium Cohort Study (MCS)  in Section \ref{ssec:teenager_multiplicity}.  The MCS is a 
survey tracking a representative sample of children born in the UK around year 2000 through their life. The sample size is $n = 11,884$ teenagers,  with $>8,000$ complete observations (Table \ref{tab:missings}),  and features several outcome measures of mental well-being, $p_x=5$ treatments measuring the use of various technologies, and $p_z=14$ controls.
Using SCA, \cite{orben:2019} concluded that the association between technology and teenager well-being was not practically relevant. 

Here, we assess the association between each (outcome, treatment) combination
and we use the closed e-BH procedure to control the FDR,
obtaining opposite conclusions to \cite{orben:2019}.

In Section \ref{ssec:teenager_combining} we illustrate how to use e-values to easily combine evidence across hypotheses and across datasets.
First, we obtain the overall evidence that each treatment is associated with at least one outcome, based on the MCS data.
Second, we combine the evidence from the MCS study with that provided by the independent Monitoring the Future (MtF) study
\citep{mtfstudy:2025}, which was analyzed by \cite{twenge:2018}.

\subsection{Multiple treatment-outcome associations}
\label{ssec:teenager_multiplicity}

We consider $L=8$ outcomes: self-assessed depressive symptoms (Mood and Feelings Questionnaire) and self-esteem (Rosenberg scale), as well as parent-assessed total difficulties, emotional problems, conduct problems, hyperactivity/inattention, peer problems and pro-sociality (all Strengths and Difficulties Questionnaire). There are $p_x=5$ treatments: time spent watching TV, playing video games, using the internet, using social media, and owning a personal computer. 
We include $p_z=14$ controls: gender, age, BMI, (self-reported) educational motivation, mother’s ethnicity, 
(self-reported) closeness to parents, presence of natural father, time spent with primary
caretaker (PC), PC’s word activity score, PC’s employment status, own longstanding
illness, PC’s psychological distress, number of siblings and household income.

The data were prepared as described in~\cite{semken:2022}. All treatment variables but TV usage were treated as
continuous variables, normalized such that 0 means “no usage” and 1 means maximum
reportable usage. TV usage was discretized into low, medium and high because its association with most outcomes was not linear (although it was monotone). The outcomes are binary and were obtained by binning originally continuous data with validated cutoffs for abnormal behavior (\cite{semken:2022}, Section S4). 
 Table \ref{tab:missings} reports the number of individuals reporting the outcomes and that of complete observations, with no missings on the treatments nor in the controls. These missings were mostly due to control covariates, with 3,446 individuals not reporting at least one control, and only 376 failing to report a treatment. Our analysis uses the complete observations, hence our findings should be interpreted as being conditional on this population. One could consider several strategies to handle the missings and view them as a data analytic choice within the multiverse, but this is beyond our scope.  

Figure~\ref{fig:exp.mcs} shows the estimated treatment effects from logistic regressions on each outcome,
and Table~\ref{tab:mcs_evalues} those with the most significant e-values.
The intervals in the figure are 95\% e-confidence intervals based on e-variable $E^{c1}$, that is, the null treatment effect values for which $E^{c1}<1/0.05$~\cite{ramdas:2025}. In other words, an e-confidence interval for $\beta_j^{(l)}$ is the set of values $b$ such that the null hypothesis $\beta_j^{(l)}=b$ is not rejected at a 0.05 level. 
Table \ref{tab:mcs_evalues} shows that the e-values flag 6 statistically significant treatment-outcome associations, indicated by e-values $\geq 20$. In contrast, standard likelihood ratio test p-values and a 5\% cutoff lead to 12 significant findings. 
These e-values and p-values do not account for multiple hypothesis testing.

Applying the closed e-BH procedure (Section \ref{ssec:fdr_control}) to control the FDR $\leq 0.05$ gave 4 significant findings (top four rows in Table \ref{tab:mcs_evalues}).
The inverse root e-variable gave the same 4 findings, albeit with smaller e-values. Alternative, less powerful, choices of e-variables in Section~\ref{ssec:define_evariables} (universal mixture, soft-rank) led to no rejections after the closed e-BH adjustment. 
 BY yielded more rejections (7 vs. 4), consistent with our simulation results for moderately sized multiverses (also shown in Table \ref{tab:mcs_evalues}). 

Our results find a significant association between technology and teenager well-being, contrary to the original study in~\cite{orben:2019}. Specifically, a high internet usage was significantly associated with low self-esteem and depressive symptoms as self-assessed by the teenagers, and high parent-assessed peer problems.
High social media use was also significantly associated with self-assessed depressive symptoms.
We remark that the odds-ratio for these associations were in the range $[2,4]$, 
that is these findings are not only statistically significant but also practically relevant.
 The main reason why our findings differ from \cite{orben:2019} is that the latter used SCA to estimate a median treatment effect across all outcomes and treatments (and across all control covariate configurations). Figure \ref{fig:exp.mcs} shows that there is significant heterogeneity in these effects. The point estimate for the association between TV and most outcomes was negative, 
whereas for social media and internet use the estimates were positive for self-assessed outcomes like depression and low self-esteem and negative for parent-assessed outcomes like high peer problems and low pro-sociality. The median of these heterogeneous effects is close to 0, which led \cite{orben:2019} to conclude that the association was too small to warrant policy change. We argue that in the presence of strong treatment effect heterogeneity, summaries such as the median can mask the presence of important individual treatment effects.
In fact, there are observational studies \cite{valkenburg:2022}, randomized controlled trials \cite{allcott:2020,allcott:2022} and natural experiments \cite{braghieri:2022} suggesting that there is a negative association between social media and mental well-being, aligning with our findings.

\subsection{Combining evidence}
\label{ssec:teenager_combining}

\begin{table}[ht]
\caption{ Combining e-values. Evidence against the global null hypothesis that each treatment has no effect on any outcome in the MCS data, and combining evidence on treatment effects on depression between the MCS and MtF studies (average calibrated e-variable $E^{c1}$) }
\label{tab:mcs_overalltest}%
\begin{tabular}{@{}lllc@{}}
\toprule
Dataset & Outcome &  Treatment    &     e-value \\
\midrule
MCS  & All 8 MCS outcomes & Internet  &  4869.0  \\ 
MCS  & All 8 MCS outcomes & Social media    &   236.0  \\ 
MCS  & All 8 MCS outcomes & Electronic games&    13.6  \\ 
MCS  & All 8 MCS outcomes & TV              &     1.17 \\ 
MCS  & All 8 MCS outcomes & Computer    &     1.02 \\ 
MCS + MTF & Depressed (adolescent) & Social media & 1.99$\times 10^{16}$  \\ 
MCS + MTF & Depressed (adolescent) & TV &         101,186        \\ 
\botrule
\end{tabular}
\end{table}

To illustrate that e-variables allow combining evidence in a simple and intuitive manner, we perform two analyses.
The first analysis is reported in rows 1-5 of Table \ref{tab:mcs_overalltest}. For each treatment, it shows an e-variable that provides evidence against the global null hypothesis that the treatment has no effect on any of the outcomes in the MCS study. This e-variable is obtained as the average of the e-variables across the 8 outcomes. There is strong evidence that internet and social media use have an effect, for electronic games there is moderate evidence and for TV and computer use there is little evidence.
One could also use Bonferroni-corrected p-values to test these global null hypotheses, controlling type I errors under arbitrary dependences between the hypotheses. The Bonferroni correction obtains statistically significant evidence (at the 0.05 level) for internet, social media and electronic games, providing a similar message to that in Table \ref{tab:mcs_overalltest}. 
Here, e-value aggregation is mainly appealing for the interpretability of the overall measure of evidence, rather than for its statistical power.

The second analysis is shown in the last 2 rows of Table \ref{tab:mcs_overalltest}. It illustrates that, after seeing the MCS data one can obtain further evidence by gathering more data, and easily aggregate that evidence, while controlling type I errors for each individual hypothesis. Specifically, \cite{twenge:2018} used the MtF data to study the association between depression (outcome) and social media and TV (treatments) controlling for gender, race, socio-economic status, grade and place of residence. Since MCS and MtF are independent studies from different countries, we obtain a combined e-value for each outcome-treatment pair present in both datasets as the product of the e-values in the MCS study (Table \ref{tab:mcs_evalues}) and that in the MtF study. 
The combined e-value is valid under the joint null that there is no association in the MCS study, conditional on the study controls, and that the same holds in the MtF study. That is, a large combined e-value provides evidence that at least one of the two associations exists in the corresponding target population. This does not imply that the association exists in both studies.

Table 2 in \cite{twenge:2018} reports a significant association between depression and social media and TV (p-values $6.69 \times 10^{-17}$ and $2.58 \times 10^{-8}$ respectively), using $n=77,487$ individuals.
The corresponding mixture calibrator e-values are $1.08 \times 10^{13}$ for social media and $126,827$ for TV.
Taking the product with the e-values from the MCS data (1,851 and 0.798) gives aggregated e-values $1.99 \times 10^{16}$ for social media and 101,186 for TV.
Interestingly, the first MCS dataset did not provide significant evidence for TV and, after failing to reject that null hypothesis with a p-value, one cannot gather more data while controlling type I errors. E-values control type I error in such sequential settings.




\section{Discussion}
\label{sec:discussion}

Multiverse analysis is a useful tool to descriptively assess the association between different treatments, outcomes for various sub-populations of individuals, control covariates, or other analysis choices, and flag whether treatment-outcome associations are robust across analysis specifications or not.
A more contentious issue is when the multiverse is used for formal statistical inference. Then, one must be careful to assess properly defined statistical estimands and to account for possible treatment effect heterogeneity. In particular, we do not recommend reporting averaged treatment effects across all analysis specifications, except for specific settings where the treatment and specifications are viewed as exchangeable.
Instead, we recommend reporting treatment effects for all analysis specifications separately and accounting for the multiple comparisons being performed. This ensures statistical validity but has a cost in statistical power, which can be a practical issue when either the sample size or the treatment effect are relatively small.
For example, in the teenager data a previous analysis had found that the average effect across all treatments and outcomes was practically irrelevant, whereas assessing each treatment-outcome association separately reveals practically-relevant associations.

 We illustrated that e-values provide adequate FDR control which, combined with the possibility of aggregating evidence across multiple hypotheses or over time and multiple studies, may provide valuable flexibility for multiverse analyses. The latter point on flexibility may be the most relevant feature, given that BY adjustments also provided FDR control at a somewhat higher power (except in large multiverses where many tests are performed). 
The concern is that e-values can have low statistical power to detect treatment effects when they truly exist. Our simulations show that this issue is much more marked for some e-values than for others, and these findings were confirmed in the teenager data. Mixture and inverse root calibrated e-variables worked best in our experiments, whereas universal mixture e-variables were extremely conservative. 
Permutation-based soft-rank e-variables were less powerful than the calibrated e-variables, particularly for small $n$, but they may be appealing in settings where the model is misspecified.

Our results suggest that e-variables can be a useful complement to multiverse analyses, provided that one wishes to perform hypothesis tests. However, they also point to significant limitations in statistical power. E-values are best suited for moderately large sample sizes and treatment effect sizes. We also remark that e-values are a recent framework and that there is active research in designing more powerful procedures, which could improve their properties when used within multiverse analyses.

\backmatter

%
%

\section*{Declarations}

The authors thank Christoph Semken for sharing the data and scripts for the teenager well-being example.

\bmhead{Funding}

DR was funded by grant 0045 within the ICREA Academia program from AGAUR (Generalitat de Catalunya), and by grant PID2022-138268NB-I00 by Ministerio de Ciencia e Innovaci\'on (MICIN)/Agencia Espa\~{n}ola de Investigaci\'on(AEI)/10.13039/501100011033 /Fondo Europeo de Desarrollo Regional (FEDER). Co-funded by the European Union (ERC, BigBayesUQ, project number: 101041064). Views and opinions expressed are however those of the author(s) only and do not necessarily reflect those of the European Union or the European Research Council. Neither the European Union nor the granting authority can be held responsible for them.

\bmhead{Ethics approval}

Not applicable.

\bmhead{Consent}

Not applicable.

\bmhead{Data}

The data are available on the \href{https://github.com/paulrognonvael/multiverse.evalues}{Github repository}
(https://github.com/paulrognonvael/multiverse.evalues).

\bmhead{Code availability}

The code is available on the \href{https://github.com/paulrognonvael/multiverse.evalues}{Github repository}
(https://github.com/paulrognonvael/multiverse.evalues).

\bmhead{Authors' contribution statements}

All authors contributed equally to this work.





\bibliography{references}

\end{document}